\documentclass[twocolumn,showpacs,showkeys,amssymb,nobibnotes,superscriptaddress,aps,prx,bibnotes]{revtex4-2}

\usepackage{graphicx}
\usepackage{dcolumn}
\usepackage{bm}
\usepackage[T1]{fontenc}
\usepackage{mathptmx}
\usepackage{mathrsfs}
\usepackage{physics}
\usepackage{simplewick}
\usepackage{amsthm,amsmath,amssymb}
\usepackage[bottom]{footmisc}

\usepackage{amssymb}
\usepackage{amsmath}
\usepackage{amsfonts}
\usepackage{bm}
\usepackage{graphicx}
\usepackage{subfigure}
\usepackage[dvipsnames]{xcolor}
\usepackage{calc}
\usepackage{epsfig}
\usepackage{color}
\usepackage{url}
\usepackage[colorlinks,citecolor=blue]{hyperref}
\begin{document}
\bibliographystyle{apsrev4-1}
\newcommand{\be}{\begin{equation}}
\newcommand{\ee}{\end{equation}}
\newcommand{\bs}{\begin{split}}
\newcommand{\es}{\end{split}}
\newcommand{\R}[1]{\textcolor{red}{YM: #1}}
\newcommand{\B}[1]{\textcolor{blue}{#1}}

\title{Updating the constraint on the quantum collapse models via kilogram masses}

\author{Qi Dai}
\affiliation{Center for Gravitational Experiment, Hubei Key Laboratory of Gravitation and Quantum Physics, School of Physics, Huazhong University of Science and Technology, Wuhan, 430074, China}
\author{Haixing Miao}
\affiliation{State Key Laboratory of Low Dimensional Quantum Physics,  Frontier Science Center for Quantum Information, Department of Physics, Tsinghua University, Beijing 100084, China}
\author{Yiqiu Ma}
\affiliation{Center for Gravitational Experiment, Hubei Key Laboratory of Gravitation and Quantum Physics, School of Physics, Huazhong University of Science and Technology, Wuhan, 430074, China}

\date{\today}

\begin{abstract}
Quantum mechanics, which governs all microscopic phenomena, encounters challenges when applied to macroscopic objects that exhibit classical behavior. To address this micro-macro disparity, collapse models such as the Continuous Spontaneous Localization (CSL) and Diosi-Penrose (DP) models have been proposed. These models phenomenologically modify quantum theory to reconcile its predictions with the observed classical behavior of macroscopic systems. Based on previous works\,([Phys.\,Rev\,D,\,95(8):084054\,(2017)] and [Phys.\,Rev.\,D,\,94:124036,\,(2016)]), an improved bound on the collapse model parameters is given using the updated acceleration noise data released from LISA Pathfinder\,([Phys.\,Rev.\,D, 110(4):042004,\,(2024)]). The CSL collapse rate is bounded to be at most $\lambda_{\rm CSL} \leq 8.3\times 10^{-11}$\,$s^{-1}$ at the mili-Hertz band when $r_{\rm CSL}=10^{-7}\,{\rm m}$, and the DP model's regularization cut-off scale is constraint to be $\sigma_{\rm DP}\sim 285.5$\,fm.  Furthermore, we discuss the potential advantages of using deep-underground laboratories to test these quantum collapse models.  Our results show the quiet seismic condition of the current deep-underground laboratory has the potential to further constrain the CSL collapse model to $\lambda_{\rm CSL}\leq3\times 10^{-11}\,{\rm s}^{-1}$ when $r_{\rm CSL}=10^{-7}\,{\rm m}$.
\end{abstract}

\maketitle

\section{Introduction}
Quantum mechanics, developed to elucidate the physical principles governing the microscopic world, faces challenges when applied to macroscopic phenomena. For instance, while quantum mechanics is considered the fundamental law of physics, macroscopic objects in our physical world are observed to behave classically\,\cite{Schrodinger1935}. This discrepancy between quantum theory and classical observations becomes particularly evident during the measurement process of a quantum system, where the wave function of the microscopic quantum system collapses upon interacting with a macroscopic apparatus.

Collapse models are phenomenological modifications of quantum mechanics proposed to solve these problems.
The most typical collapse models include the continuous spontaneous localization\,(CSL) model\,\cite{ghirardi1990markov}, where the quantum state follows a continuous stochastic evolution in its Hilbert space with a suppression of the decoherence;  and the Diosi-Penrose\,(DP) model\,\cite{diosi1989models,penrose1996gravity}, where stochastic fluctuation of the gravitational field drives the stochastic quantum state evolution. 

The CSL model has two parameters $\lambda_{\rm CSL}$ and $r_{\rm CSL}$:
the $\lambda_{\text{CSL}}$ is a rate that spontaneous collapse acts on a single nucleon and $r_{\text{CSL}}$ is the length scale, beyond which the collapse effect is significantly enhanced. The DP model is characterized by a cut-off length scale $\sigma_{\rm DP}$ which is introduced to regularize the divergence due to the point-particle approximation of the ions. These two models share a similar density matrix evolution equation presented as:
\begin{equation}
    \dfrac{d}{dt} \hat{ \rho }_{t} = -\dfrac{i}{\hbar}\,\left[\hat{H},\hat{ \rho }_{t}\right] +  \iint d^3\mathbf{x} d^3 \mathbf{y} D(\mathbf{x-y})[\hat{M} (\mathbf{x}), [\hat{M} (\mathbf y), \hat{ \rho }_{t}]]\label{EoM_for_2},
\end{equation}
with different correlator $D(\mathbf{x-y})$\,\cite{carlesso2022present}.
For the CSL model, the correlator is 
\begin{equation}
   D_{\rm CSL}(\mathbf{x-y})= \dfrac{\lambda_{\rm CSL}}{m_{0}^2}\text{exp}\left[-\dfrac{\mathbf{(x-y)}^2}{4r^2_{\rm CSL}}\right],
\end{equation}
where $m_0$ is the mass of the mass of a nucleon, $\lambda_{\rm CSL}$ is a rate that spontaneous collapse acts on a single nucleon and $r_{\rm CSL}$ is the length scale, beyond which the collapse effect is significantly enhanced.
 Ghirardi, Rimini and Weber\,\cite{GRW_ghirardi1986unified}, proposed $r_{\rm CSL} = 10^{-7}\,{\rm m}$ and $\lambda_{\rm CSL} = 10^{-16}\,{\rm s}^{-1}$ with the consideration that superposition of macroscopic objects will collapse in $10^{-7}$ second; and Adler, after estimating time for a latent image formation
and heating of the intergalactic medium\,\cite{adler2007lower}
gives $\lambda_{\rm CSL} = 10^{-8\pm2}\,{\rm s}^{-1}$ when $r_{\rm CSL} = 10^{-7}\,{\rm m}$ and $\lambda_{\rm CSL} = 10^{-6\pm2}\,{\rm s}^{-1}$ when $r_{\rm CSL} = 10^{-6}\,{\rm m}$. 
As for the DP model, 
\begin{equation}
    D_{\rm DP}(\mathbf{x-y}) = \dfrac{G}{\hbar}\dfrac{1}{\mathbf{|x-y|}},
\end{equation}
with $G$ the gravitational constant, $\hbar$ the reduced Planck constant. The $\hat{M} (\mathbf y)$ term is a mass density operator, added to involve gravity.
 
 Ever since the model was proposed, people have been running experiments\,\cite{carlesso2022present,altamura2024noninterferometric,donadi2021underground} and developing sophisticated techniques \,\cite{marchese2023optomechanics,zicari2024criticality} to constrain its parameters. Due to their extreme measurement accuracy, gravitational wave detectors are promising candidates for testing fundamental physics, for example, AURIGA\,\cite{carlesso2018non}, LIGO\,\cite{Carlesso_GWD_PhysRevD.94.124036} and LISA Pathfinder\,\cite{helou2017lisa}. 
 
 
Nimmrichter \textit{et al.}\,\cite{nimmrichter2014optomechanical} applied the CSL model to the macroscopic test mass in an optomechanical system, and obtained the correlation function of the effective force noise exerting on the test mass:
\begin{equation}
\langle F_{\text{CSL}}(t)F_{\text{CSL}}(t+\tau)\rangle  =D_{\text{CSL}}\delta(\tau) ,
\end{equation}
where
\begin{equation}
D_{\text{CSL}}=\lambda_{\text{CSL}}\left[\dfrac{\hbar}{r_{\text{CSL}} }\right]^2 \alpha,
\end{equation} 
with $\alpha$ a geometry factor of the form
\begin{equation}
\alpha \approx\dfrac{8\pi\rho^2 r_{\text{CSL} }^4} {b^2m^2_0},
\end{equation}
They pointed out that the white component of the noise spectral density contains at least three parts, the CSL effective noise, the thermal Brownian noise, and measurement shot noise, namely
\begin{equation}
S_{F}(f) = D_{\text{CSL}}+D_{k_{B}T}+D_{\rm measure}+\text{other noises}.
\end{equation}
Furthermore, quantum mechanics imposes a sensitivity limit (standard quantum limit, or SQL) for measuring the force acting on a free mass test mass $S^{\rm SQL}_{ FF}= \hbar M (2\pi f)^2$. Comparing this sensitivity limitation to the CSL effective noise, Nimmrichter\,\cite{nimmrichter2014optomechanical} concludes that low-frequency measurement is favored in testing the collapse model.

Similarly, applying the Di{\'o}si-Penrose model\,\cite{diosi1989models} to the macroscopic test mass would lead to a similar noise while the $D_{\text{DP}}$ is of the form:
\begin{equation}
    D_{\mathrm{DP}} \approx \frac{G \hbar}{6 \sqrt{\pi}}\left(\frac{a}{\sigma_{\mathrm{DP}}}\right)^{3} M \rho,
\end{equation}
where $M, a,\rho$ are the mass, crystal lattice constant, and mass density of the test mass. The $\sigma_{\rm DP}$ is the regularization length scale parameter introduce to regularize the divergence due to the point mass approximation.
  
\section{Updated constraint of collapse models by LISA pathfinder}
  
LISA Pathfinder\,(LPF) is a mission aiming to demonstrate the technology required by space-borne gravitational wave detectors, in particular, to assess the accuracy of the achievable free-fall condition. The core of LPF consists of a pair of test masses and the optical readout system targeted at measuring their relative acceleration noise, which is denoted as $\Delta g$. Importantly, the two test masses in the LPF are almost free-falling masses with very soft rigidity due to the interaction with the capacity sensors, and the frequency band is on the mili to sub-mili Hertz, which satisfies the Nimmrichter's low-frequency condition. Therefore, LPF and similar devices can be used to test the collapse model.
  
In 2016, Helou\textit{ et al.}\cite{helou2017lisa,Carlesso_GWD_PhysRevD.94.124036} use LPF's data to constrain the CSL model. Their idea can be briefly introduced as follows.
 Since LPF measures the relative acceleration between two test masses, the acceleration noise spectrum can be related to the thermal noise spectrum by
\begin{equation}
 	S_{a}(f) =4S_{F}(f)/M^2,
\end{equation}
which relates to the CSL model as
\begin{equation}
 	\lambda_{\text{CSL}}=\dfrac{m^{2}_{0}}{32\pi\hbar^2r_{\text{CSL}}^{2} } \left(\dfrac{M}{\rho}\right)^2\dfrac{1}{b^2} S_a.
\end{equation}
Using parameters $\rho =19881\,\text{kg}/{\rm m}^3$, $M = 1.928\,\text{kg}$, $b=46\,\text{mm}$, they come to $\lambda_{\rm max} = 2.96\times10^{-8}\,{\rm s}^{-1}$ when $r_{\rm CSL}=10^{-7}\,{\rm m}$.
 
Similarly, the bound on the DP model can be calculated as:
 \begin{equation}
    \sigma_{\mathrm{DP}}^{\min }=a\left(\frac{2 \hbar G}{3 \sqrt{\pi}} \frac{\rho}{m} \frac{1}{S_{a}}\right)^{1 / 3}=40.1 \mathrm{fm},
\end{equation}
where $a=4.0\,\mathring{\rm A}$ is the lattice constant for LPF's test masses.

They use the data on the first 55 days of science operations\cite{LPF_run_PhysRevLett.116.231101}, namely
\begin{equation}
	\sqrt{S_a} = 5.2\,\text{fm}\cdot \text{s}^{-2}/\sqrt{\text{Hz}},
\end{equation}
which is a conservative one, as they have assumed that all measured noise is attributed to CSL.
 
\begin{figure}[h]
		\includegraphics[width=0.45\textwidth]{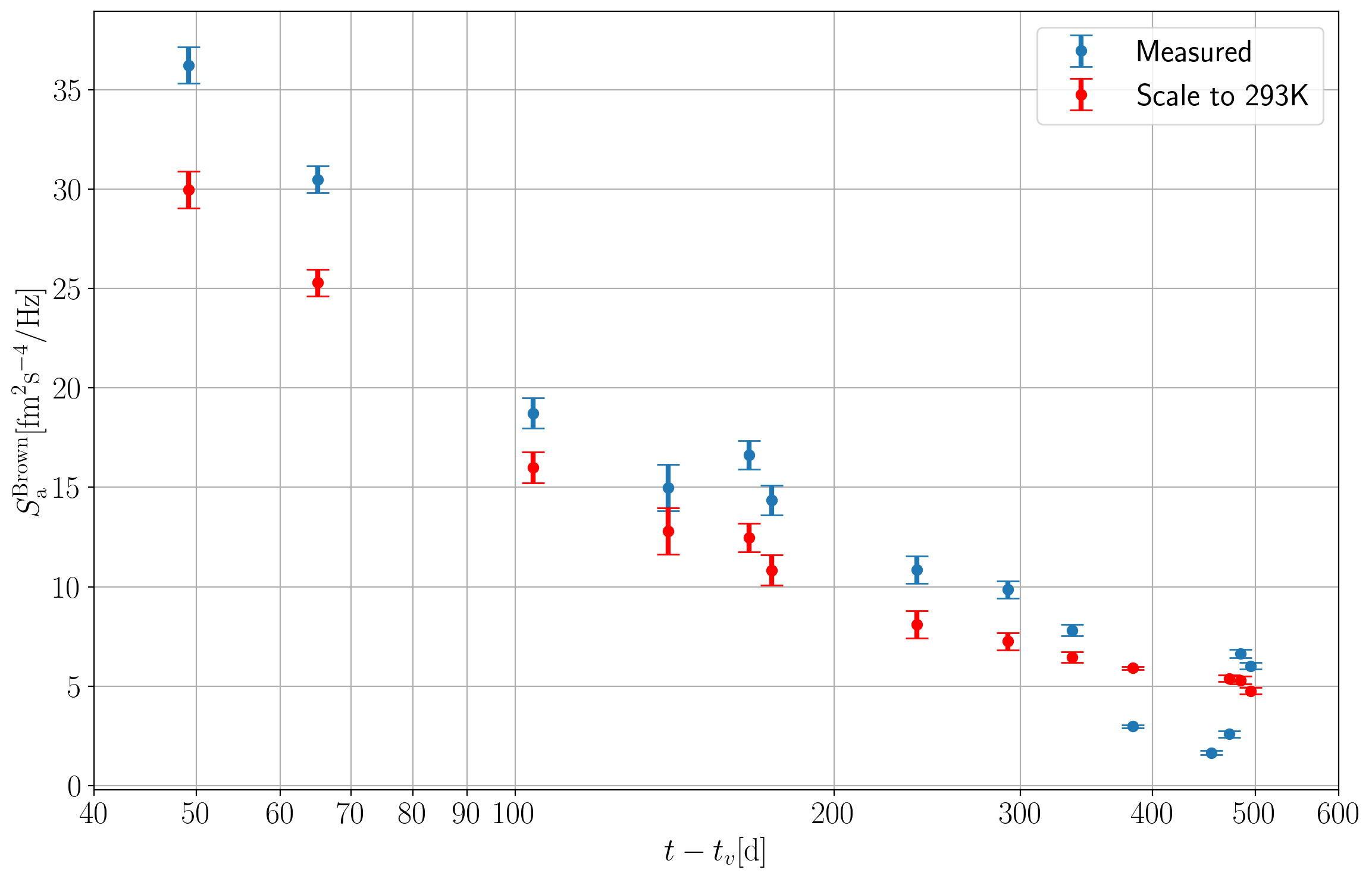}
	\caption{Brownian noise level of the LPF reproduced from Ref.\,\cite{armano2024depth} with a linear-scale vertical axis. Blue points: the measured Brownian noise level with 1-$\sigma$ uncertainty from ten science operation runs. Red points: rescaled data using the model in Ref.\,\cite{armano2024depth}. We use the Brownian for the 10th science operation run in February 2017 at the temperature: 284.72$\pm$0.03\,K.}
	\label{excess_bwnoi}
\end{figure}

Now LISA pathfinder group has performed an in-depth analysis and 
 the total acceleration noise is broken down into noise contribution from different sources\,\cite{armano2024depth}. Their motivation is to analyze the various environmental noise source of test masses and to find a physical explanation for the $f^{-1}$ tail at $f<3$\,mHz.  
Essentially, the measured acceleration noise consists of a white noise component and a colored component as:
\be
S_{a}(f)=S^{\rm Brown}_{a}+S^{\rm color}_{a}(f),
\ee
where the white noise component $S^{\rm Brown}_{aa}$ is attributed to Brownian noise mostly coming from the outgassing effect, while the colored component $S^{\rm color}_{aa}(f)$ is named as ``noise over Brownian" noise in~\cite{armano2024depth}. Furthermore, Ref.\cite{armano2024depth} presents the evolution of the mean value and the uncertainty of $S^{\rm Brown}_{a}$ according to the ten mission scientific operations that lasted for more than 16 months, from March 1, 2015 to July 18, 2017, which we reproduced in Fig.\ref{excess_bwnoi}.  They found long-term evolution of the PSD of Brownian noise $S^{\rm Brown}_a$ follows a power law of $t^{-0.8}$ during the whole mission. This power law evolution can be explained with a model of water diffusion in a polymer-full environment\,\cite{chiggiato2020outgassing}, which has a similar feature as the outgassing process of the water molecule released from the metal test mass.  

\begin{figure}[h]
  		\includegraphics[width=0.45\textwidth]{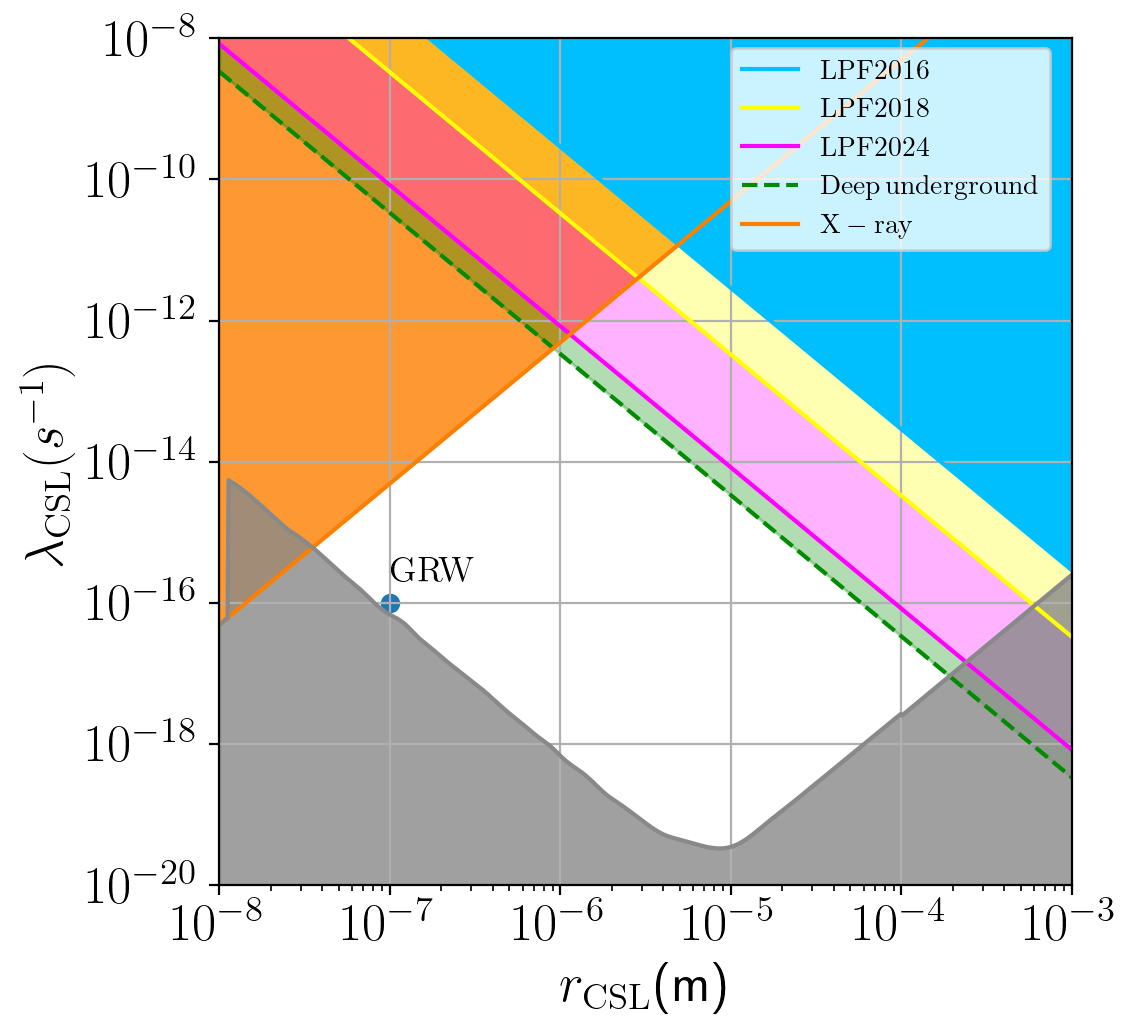}
 	\caption{Bound on the $(\lambda_{\rm CSL},r_{\rm CSL})$ parameter phase space of the CSL model\,\cite{helou2017lisa,carlesso2018non}. Blue and yellow lines are the bound based on the LPF's data in 2016\,\cite{LPF_run_PhysRevLett.116.231101}, and the LPF's best data\,\cite{armanoPhysRevLett.120.061101}. The magenta line represents the bound derived by the updated LPF noise data in\,\cite{armano2024depth}, while the dashed line represents the bound derived by the sensitivity of the proposed deep-underground device. The grey area is excluded from the requirement that macroscopic superpositions do not persist in time.}
 	\label{fig:CSL_bound}
 \end{figure}

Since the CSL force noise is white, it will also be included in the Brownian component $S^{\rm Brown}_{aa}$.  Therefore, the uncertainty of the Brownian noise spectrum can be used to constrain the CSL model.  Note that the CSL force noise is a white noise with a spectrum \emph{independent} of the gas pressure and temperature. Thus, the CSL force noise can be constrained by the minimum uncertainty of the Brownian noise among all 10 science operations.  With the minimum uncertainty $\sigma_{\rm Brown}=\Delta S^{\rm Brown}_{a} = 0.075\,{\rm fm}^2{\rm s^{-4}}/{\rm Hz}$ for the run in February 2017, the $\lambda $ in CSL model will be constrained by
 \begin{equation}
 	\lambda_{\text{CSL}} < 8.3\times10^{-11} {\rm s^{-1}},
 \end{equation}
 when $r_{\rm CSL}=10^{-7}\,{\rm m}$, and the bound on the $(\lambda_{\rm CSL}, r_{\rm CSL})$ parameter phase space is shown as the magenta line in Fig.\,\ref{fig:CSL_bound}.
 For the DP model, we can update $\sigma_{\text{DP}}$ with the new bound
 \begin{equation}
     \sigma_{\text{DP}}>285.5\,\text{fm},
 \end{equation}
 however the strongest bound for DP model is set by X-ray emission test\,\cite{arnquist2022search}, at $\sigma_{\text{DP}}>4.94\times10^{5}$\,fm, which is much larger than the bound derived from macroscopic test masses. However, the kilogram test masses, due to its macroscopicity, is still a platform is still an important system that worth the experimentally tested for complementary reasons.

  \section{Constraining collapse models in deep underground laboratory}
  The above constraint on the collapse model is based on the linear relative acceleration between two test masses. Moreover, Ref.\,\cite{carlesso2018non} also pointed out that a rotation optomechanical device can also be used to constrain the CSL model. An analysis of the possible rotational noise from LISA Pathfinder was also presented in\,\cite{carlesso2018non}, where the bound can be improved.  However, additional degradation exists due to the complicated environmental effects, such as imperfect cancelation of the actuation noise for spaceborne experiments and the seismic noise for ground-based experiments.  Therefore, more delicate experiments designed to measure rotation directly are needed. This section is devoted to investigate the potential of the deep underground laboratory to constraint the collapse model.
  
  \subsection{Advantages of deep underground laboratory}
Deep underground laboratories, such as Gran Sasso National Laboratory\,(LNGS),  Sudbury Neutrino Observatory\,(SNO), and Kamioka Mine Underground Laboratory\,(KMSSL)\,\cite{UGL_bettini2008underground,UGL_smith2012development}, Jinping Underground Laboratory\,(JUL)\cite{UGL_kang2010status,UGL_li2015second}, are designed to study extremely weak interaction process of particle physics, such as dark matter collision with standard model particles, neutrino physics, or double-$\beta$-decay. These deep underground sites have the following advantages for testing the collapse model. 

\emph{Cosmic ray shielding} --- The Earth's surface is constantly bombarded by high-energy cosmic rays from outer space, which can charge the test masses\,(or its metal surface) and introduce additional noises. Deep underground laboratories, being shielded by thick rock layers, can significantly block these cosmic rays, thereby reducing background noise and benefiting the detection of very faint signals. This is crucial for experiments like dark matter searches, where the interaction of dark matter with ordinary matter is extremely weak and can only be detected in environments with minimal background noise. 

\emph{Thermal Stability} --- Temperature variation at low frequencies can significantly affect the sensing of the test mass motion via thermoelastic expansion. The temperature in underground environments varies less than on the surface, which helps maintain experimental equipment's stability, thereby increasing its sensitivity. Furthermore, the thermal stability issue can be solved using a careful common-mode rejection design as we shall detail later.

\emph{Seismic isolation} --- The underground laboratory is known to have low ambient seismic noises, due to the thick rock layers's attenuation of the seismic wave and the more stable geophysical structures. A better seismic environment would be helpful for the experiment targeted at high-precision displacement sensing. For the optomechanical device that is targeted at testing the collapse models, the displacement seismic noise could be significantly reduced by common-rejection schemes, while the rotational seismic noise will dominate. 

\emph{Low magnetic field} --- The underground laboratory has a low magnetic field feature. An alloy test mass or the metal-coated surface of the test mass has magnetic susceptibility. The fluctuating magnetic field would exert a stochastic Ampere force on the test mass, disturbing the test masses' displacements. While this problem can also be solved by choosing the appropriate test mass substrate and coating material, or by a carefully designed electromagnetic shielding cage, the low magnetic environment is still beneficial. 

\subsection{A proposed deep-underground optomechanical device}

Compared with the space mission for testing the CSL model, the ground-based experiment with more controllability has advantages despite the noisy environment.  For example, the ground-based environment can be carefully sensed so that various components of the noise budget can be calibrated with a high precision.
\begin{table}[h]
\caption{\label{tab:sampling_parameters}%
Parameters and the noise models for the experiment }
\begin{ruledtabular}
\begin{tabular}{ccc}
Parameter&Description&Value\\
\hline
$M$ & Mass of a cube &1\,kg\\
$\rho$ & Mass density of fused silica & 2.33\,g\,${\rm cm}^{-3}$ \\
$L$ & Torsional pendulum arm length & 0.1\,${\rm m}$ \\
$\omega_m/(2\pi)$ & Torsional  frequency & 1\,${\rm mHz}$ \\
$Q$ & Vacuum quality factor & $10^6$ \\
$T$ & Temperature & 300\,${\rm K}$ \\
$\Delta T $ & Temperature stability & $10^{-4}\,{\rm K}/\sqrt{\rm Hz} $ \\
$p_{\rm air}$ & Air pressure & $10^{-7}\,{\rm Pa}$\\
$\lambda$& Optical wavelength& $1064\,{\rm nm}$\\
$P_l$ & Laser power & 1\,${\rm mW}$ \\
$\delta I/I$ & Relative laser intensity noise & $10^{-5}(\frac{1\rm mHz}{f})/\sqrt{\rm Hz}$\\
$\delta f$ & Laser frequency noise & $10^{4}(\frac{1\rm mHz}{f}){\rm Hz}/\sqrt{\rm Hz} $\\
$\alpha_E$& Thermo expansion coefficient&$ 5.5\times 10^{-7}/{\rm K}$\\
\end{tabular}
\end{ruledtabular}
\end{table}

\begin{figure*}
 		\includegraphics[width=1.0\textwidth]{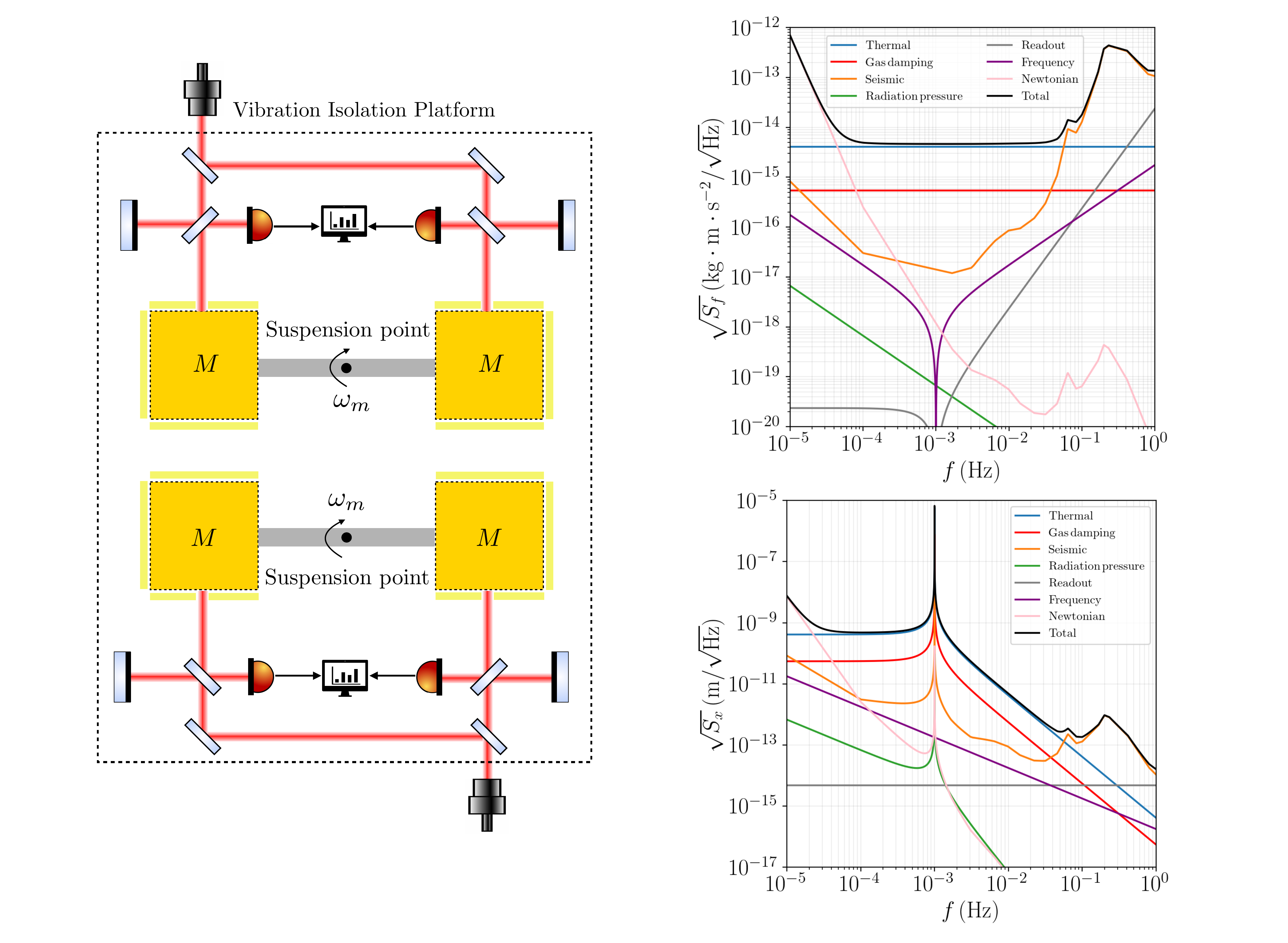}
 	\caption{Left panel: Schematic diagram of the optomechanical device in the deep underground laboratory to test collapse model.  The two interferometers sense the displacement motion of the two test masses, where the difference between the two interferometric sensing data gives the rotation information of the test masses. Two such optomechanical torsion balance is installed in the same platform, and the results of the measurement from these two torsion balance is combined to further reduce the seismic effect and Newtonian noise. The two optical fields are phase locked  Right panel: Force (upper-right panel) and displacement noise (lower-right panel) budget of the proposed optomechanical device. The noise floor is dominated by the Brownian thermal noise (including the gas damping) of the torsional balance and the rotational seismic noise. The thermal noise can be in principle calibrated, and the seismic noises dominate the sensitivity, which is used to constrain the collapse model.}
 	\label{fig:scheme_sensitivity}
 \end{figure*}

Our proposed optomechanical device is a test-bed platform for demonstrating the key technology of the spaceborne gravitational wave detector, e.g. located at Jinping, Sichuan, China. The schematic diagram of the device is shown in Fig.\,\ref{fig:scheme_sensitivity}, with key features summarized as follows. 

Two 1\,kg test masses made by fused silica are connected with a fused silica arm and suspended, forming a low-frequency torsion pendulum with resonant frequency at 1\,mHz and quality factor $Q=10^6$. The fused silica cubics are coated with gold and surrounded by capacity sensors, which are used to suppress the long-term drifting of the test mass displacement. The charging rate of the metal surface of the deep-underground system is negligibly low compared with the space environment. The test mass motion is recorded by the output phase of the two interferometers, where the angular motion is given by $\delta \theta=(x_1-x_2)/L$. 

The torsion balance system is duplicated and installed in one platform, where the laser fields coupled to these two torsion balances are phase-locked so that they share the same frequency\,(intensity noise can be suppressed individually) The power of the sensing laser is set to be 1\,mW, with the $1/f$ relative laser intensity noise and frequency noise shown in Tab.\,\ref{tab:sampling_parameters}. The system is put in a vacuum tank with the air pressure $p_{\rm air}\sim10^{-7}$\,Pa. For one torsion balance, the seismic displacement-induced effect and the thermal expansion noise can be ignored by common-mode rejection. In contrast, seismic rotation-induced noise, Newtonian noise, and torsion thermal noise (due to mechanical loss and gas damping) still matter. For example, the rotational motion will couple to the torsion pendulum through the suspension wire, of which the effect is estimated using the rotational seismic data in\,\cite{Brotzer2023}, and the Newtonian noise is estimated based on \cite{NN_saulson1984terrestrial,NN_harms2013low}.  However, since the seismic rotation-induced noise and Newtonian noise affect the two torsional balances in a common-motion way, combining the measurement results of these two torsion balances could reduce these noises.

 To present an estimation of the potential of the underground laboratory site, we summarise the sampling parameters in Tab.\,\ref{tab:sampling_parameters}, and an optimistic estimation of the noise budget is given in Fig.\,\ref{fig:scheme_sensitivity}. The dominating noise sources are thermal Brownian noise, residue gas damping noise, and rotational seismic noise, with the force and the displacement sensitivity at the level of femto-Newton and nanometers, respectively. Furthermore, since the temperature, air pressure, and the quality factor can be auxiliary measured at a high precision,  the thermal noise and residue gas damping noise can be calibrated. Therefore, the rotational seismic noise finally constrains the CSL model. As we have mentioned previously, the seismic noise force acts commonly on the two torsion balances, hence the seismic data in the measurement results of these two optomechanical system are correlated. This allow us to further reduce the seismic effect, which we assumes to be 10 percent of the seismic noise for a single torsion balance as an estimation, as shown in Fig.\,\ref{fig:scheme_sensitivity}.
 After calibrating thermal noise and gas damping noise, and optimally combining the interferometric data of the two torsional balance system, the seismic-dominated residue force noise floor could be at the level of $1\times10^{-17}\,{\rm N }/\sqrt{\rm Hz}$. This can be used as an estimation to constrain the collapse model, with the parameter bound as $\lambda_{\rm CSL}<3\times10^{-11}\,{\rm s}^{-1}$ at $r_{\rm CSL}=10^{-7}\,{\rm m}$, shown as the green dashed line in Fig.\,\ref{fig:CSL_bound}. For DP model, the regularization length scale $\sigma_{\rm DP}$ is estimated to be larger than $945.2$\,fm, using the lattice constant for fused silica $a=5.0\,\mathring{\rm A}$, which is also a stronger bound.
 
 \section{Conclusion}
Testing the possible deviation of standard quantum mechanics at the macroscopic scale becomes a interesting topic of numerous theoretical and experimental studies, especially considering the rapid progress of the development of high precision measurement technology\,\cite{Friedman2000}\,\cite{Romero-Isart2011}. Up to now, the portion of the collapse models' parameter space that has not been explicitly excluded remains vastly broad. For example, it still spans several orders of magnitude in both the correlation length $r_{\rm CSL}$ 
and the localization rate $\lambda_{\rm CSL}$. Therefore it is still an important experimental task to push forward the measurement sensitivity for testing the collapse models. In this work, an in-depth analysis of the LISA Pathfinder data for the relative acceleration noise\,\cite{armano2024depth} allows us to explore the quantum collapse models at the macroscopic mass scale\,(kilogram) and the low frequency region\,(mili-Hz to sub-mili-Hz), and update the bound on the spontaneous collapse models. The parameters of the collapse models are further constraint the $\lambda_{\rm CSL}$ to be smaller than $8.3\times 10^{-11}\,{\rm s}^{-1}$ when $r_{\rm CSL}=10^{-7}\,{\rm m}$ for the CSL model, while the updated bound for the Diosi-Penrose model is $285.5\,{\rm fm}$. Moreover, we propose a dual torsional balance low frequency optomechanical device, combining the advantages of the deep-underground laboratories, to test the quantum collapse model. The potential of such a proposed device is investigated, and we found that the CSL collapse model can be further constrained to $\lambda_{\rm CSL}<3\times 10^{-11}\,{\rm s}^{-1}$ at $r_{\rm CSL}=10^{-7}\,{\rm m}$, and the DP model can be constrained to $\sigma_{\rm DP}>945.2\,$fm. These investigations show that testing macroscopic quantum mechanics using deep-underground laboratory site could be a promising future direction.

\acknowledgements
Y.M. thanks Professor Shun Wang for helpful discussions on the test mass materials. Y.M. is supported by the National Key R$\&$D Program of China ``Gravitational Wave Detection" (Grant No.: 2023YFC2205801), and National Natural Science Foundation of China under Grant No.12474481. H.M. would like to acknowledge the support from the National Key R$\&$D Program of China ``Gravitational Wave Detection" (Grant No.: 2023YFC2205800), and support from Frontier Science Center for Quantum Information,.

\bibliographystyle{unsrt}
\bibliography{ref_article}
\end{document}